\documentstyle[11pt, aasms4]{article}
\input epsf
\tighten
\begin{document}
\def\plotone#1{\centering \leavevmode
\epsfxsize=\columnwidth \epsfbox{#1}}

\topmargin 0.0cm
\oddsidemargin 0.2cm
\textwidth 16cm 
\textheight 21cm
\footskip 1.0cm


\newenvironment{sciabstract}{%
\begin{quote} \bf}
{\end{quote}}


\renewcommand\refname{References and Notes}


\newcounter{lastnote}
\newenvironment{scilastnote}{%
\setcounter{lastnote}{\value{enumiv}}%
\addtocounter{lastnote}{+1}%
\begin{list}%
{\arabic{lastnote}.}
{\setlength{\leftmargin}{.22in}}
{\setlength{\labelsep}{.5em}}}
{\end{list}}

\def\plotone#1{\centering \leavevmode
\epsfxsize=\columnwidth \epsfbox{#1}}

\def\wisk#1{\ifmmode{#1}\else{$#1$}\fi}

\def\lt     {\wisk{<}}
\def\gt     {\wisk{>}}
\def\le     {\wisk{_<\atop^=}}
\def\ge     {\wisk{_>\atop^=}}
\def\lsim   {\wisk{_<\atop^{\sim}}}
\def\gsim   {\wisk{_>\atop^{\sim}}}
\def\kms    {\wisk{{\rm ~km~s^{-1}}}}
\def\Lsun   {\wisk{{\rm L_\odot}}}
\def\Zsun   {\wisk{{\rm Z_\odot}}}
\def\Msun   {\wisk{{\rm M_\odot}}}
\def\um     {$\mu$m}
\def\mic     {\mu{\rm m}}
\def\sig    {\wisk{\sigma}}
\def\etal   {{\sl et~al.\ }}
\def\eg     {{\it e.g.\ }}
 \def\ie     {{\it i.e.\ }}
\def\bsl    {\wisk{\backslash}}
\def\by     {\wisk{\times}}
\def\half {\wisk{\frac{1}{2}}}
\def\third {\wisk{\frac{1}{3}}}
\def\nwm2sr {\wisk{\rm nW/m^2/sr\ }}
\def\nw2m4sr {\wisk{\rm nW^2/m^4/sr\ }}

\title{Detecting Population III stars through observations of near-IR cosmic infrared
background anisotropies.}

\author{
A. Kashlinsky$^{1,{a},\ast}$, R.Arendt$^{1,a}$, Jonathan P. Gardner$^{2,b}$, J. Mather$^{1,{b}}$,S. Harvey Moseley$^{1,b}$\\
{$^{1}$Laboratory for Astronomy and Solar Physics, Code 685, Goddard Space Flight Center, Greenbelt MD 20771}\\
{$^{2}$Laboratory for Astronomy and Solar Physics, Code 681, Goddard Space Flight Center, Greenbelt MD 20771}\\
{$^{a}$SSAI, $^{b}$ NASA}\\
{$^\ast$To whom correspondence should be addressed; E-mail: kashlinsky@stars.gsfc.nasa.gov}
}

\begin{abstract}
Following the successful mapping of the last scattering surface by
WMAP and balloon experiments, the epoch of the first stars, when
Population III stars formed, is emerging as the next cosmological
frontier. It is not clear what these stars' properties were, when
they formed or how long their era lasted before leading to the stars and
galaxies we see today. We show that these questions can be answered with the current and future measurements of the near-IR cosmic infrared background (CIB). Theoretical arguments suggest that Population III stars were 
very massive and short-lived stars that formed at $z\sim 10-20$ at
rare peaks of the density field in the cold-dark-matter Universe.
Because Population III stars probably formed
individually in small mini-halos, they are not directly accessible
to current telescopic studies. We show that these stars 
left a strong and measurable signature via their contribution to
the CIB anisotropies for a wide range
of their formation scenarios. The excess in the recently measured near-IR CIB
anisotropies over that from normal galaxies can be explained by contribution from early Population
III stars. These results imply that Population III were indeed very
massive stars and their epoch started at $z\sim 20$ and lasted past
$z\lsim 13$. We show the importance of accurately measuring the CIB anisotropies produced by Population III with future space-based missions.
\end{abstract}


\section{Introduction}

The cosmic infrared background (CIB) arises from the accumulated
emission of galaxy populations spanning a large range of redshifts.
The earliest epoch for the production of this background occurred
when star formation first began, and contributions to the CIB
continue through the present epoch. The CIB is thus an integrated
summary of the collective star forming events, star-burst activity,
and other luminous events in cosmic history to the present time.

With the WMAP measurements (\cite{wmap}) the structure of the last
scattering surface has been mapped and a firm cosmological model
has now emerged: the Universe is flat, dominated by the vacuum
energy or an exotic quintessence field and is consistent with the
inflationary paradigm and a cold-dark-matter (CDM) model. In this
paper, we adopt the $\Lambda$CDM model with cosmological parameters
from the WMAP and other observations: $\Omega_{\rm baryon}=0.044,
h=0.71, \Omega_{\rm m}=0.3, \Omega_\Lambda=0.7, \sigma_8=0.84$.
Following the last scattering at $z\sim 1000$, the Universe entered
an era known as the "Dark Ages", which ended with the
star formation which produced Population III stars. The WMAP
polarization results (\cite{kogut}) show that the Universe had an optical depth
since last scattering of $\tau \sim 0.2$, indicating an unexpectedly
early epoch of the first star formation ($z_* \sim 20$).
From the opposite direction in $z$, optical and IR telescopes are
now making progress into understanding the luminosity history during
the most recent epoch of the universe ($z < 5$), but the period
from recombination to the redshift of the galaxies in the Hubble
Deep Field (HDF) remains largely an unexplored era. The first
objects in the Universe probably formed in mini-halos of $\sim
10^{5-6} M_\odot$ (e.g. \cite{miralda}) (barring the possible radiative feedback effects), which would remain largely
inaccessible to current instruments and conventional methods.

A consensus based on recent numerical investigations is now emerging
that fragmentation of the first collapsing clouds at redshift $z_*$
was very inefficient so that the first metal-free stars 
were extremely massive with mass 
$\gsim 100 M_\odot$ (\cite{abel,bromm}). Such stars would live
only a few million years, which is much less than the age of the
Universe ($\simeq 2\times10^8$ years at $z=20$), making their
direct detection still more difficult. It is unclear how long the
era of Population III stars lasted and when the
Population II stars formed with metallicities $\sim 10^{-3}$ solar
(\cite{ostriker}). On the other hand the net radiation produced by
these massive stars will give substantial contributions to the
total diffuse background light (\cite{mjr}) and since their light is
red-shifted much of that contribution will be today in the infrared
bands (\cite{bond}) contributing significantly to the near-IR CIB.

Observationally, the CIB is difficult to distinguish from the
generally brighter foregrounds from the local matter within the
Solar system, and the stars and ISM of the Galaxy. A number of
investigations have attempted to extract the isotropic component
(mean level) of the CIB from ground- and satellite-based data (see
\cite{cibreview} for a recent review). These analyses of the {\it
COBE} data have revealed the CIB at far-IR wavelengths $\lambda >
100$\um (\cite{hauser}), and probably at near-IR wavelengths from 1
- 3 \um\ (\cite{dwek,gorjian,cambresy}) with additional support from
analysis of data from {\it IRTS} (\cite{irts}). However, none of the
reported detections of the isotropic CIB are very robust (especially
at near--IR wavelengths), because all are dominated by the systematic
uncertainties associated with the modeling and removal of the strong
foreground emission. Furthermore, the near-IR colors of the mean
CIB do not differ greatly from those of the foregrounds, limiting
the use of spectral information in distinguishing the true CIB from
residual Galactic or solar system emission. Because of the difficulty
of exactly accounting for the contributions of these bright
foregrounds when measuring the CIB directly, and the difficulty in
detecting {\it all} the contributing sources individually, it was proposed to
measure the structure or anisotropy of the CIB via its angular
power spectrum (\cite{paper1}). For a relatively conservative set of
assumptions about clustering of distant galaxies, fluctuations in
the brightness of the CIB have a distinct spectral and spatial
signal, and these signals can be more readily discerned than the
actual mean level of the CIB.

\begin{figure}[h]
\centering
\leavevmode
\epsfxsize=0.5
\columnwidth
\epsfbox{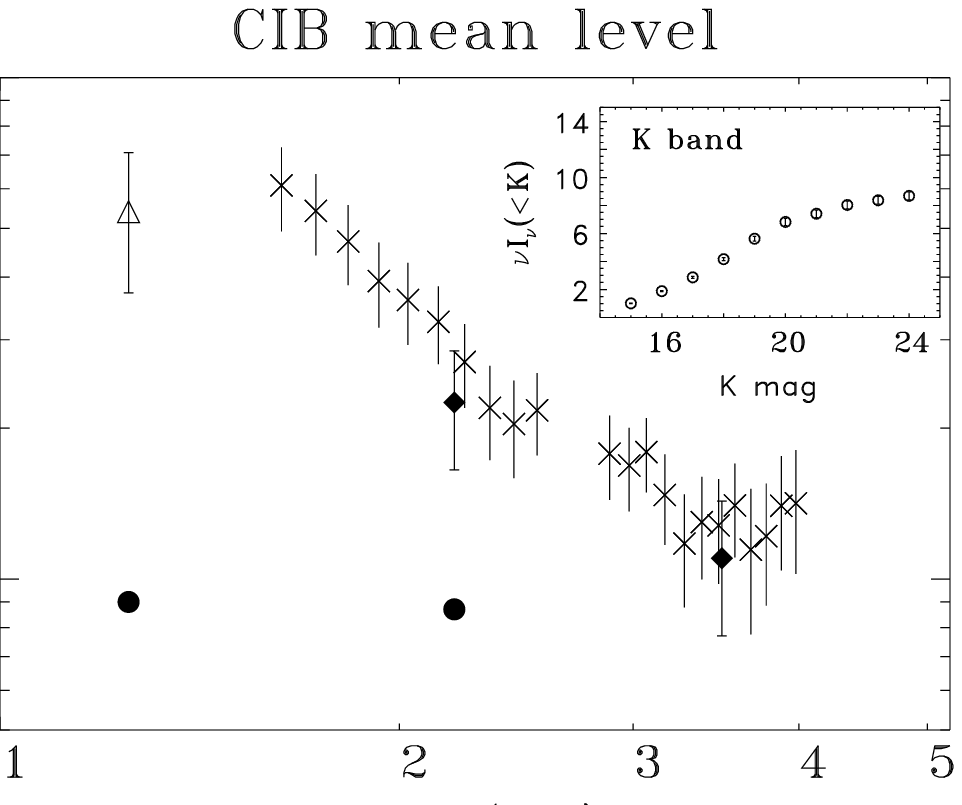}
\vspace{0.5cm}
\caption[]{
\scriptsize{
Mean levels of the near--IR CIB from IRTS (crosses from \cite{irts}
and from COBE DIRBE (filled diamonds - at 2.2 \um\ from \cite{gorjian}
and at 3.5 \um\ from \cite{dwek}. Cumulative flux from {\it observed}
J and K band galaxies is shown with the filled circles
(\cite{cambresy,gardner,koscience}) Open triangle shows the CIB
results of the DIRBE J-band analysis from \cite{cambresy}. The
insert shows the cumulative flux distribution in \nwm2sr as a function
of K magnitude from galaxies from deep galaxy surveys (\cite{totani}).
}
}
\label{cib_dc}
\end{figure}

In Fig. \ref{cib_dc}, we summarize the current near-IR CIB measurements
and compare the measured CIB at 1.25 and 2.2 \um\ to the total flux
from the observed galaxy populations. There is a statistically
significant excess over the CIB expected from normal galaxies; e.g.
at 1.25 \um\ (J band) and 2.2 \um\ (K band), while an integration
of the measured galaxy counts gives only 30-40\% of the observed
CIB (\cite{koscience,gardner,madau}). If this deficit CIB does not
come from undetected low surface brightness emission associated
with local galaxies, it must come from fainter galaxies at high
$z$.

The measurements of CIB structure or fluctuations lead to similar
conclusions (\cite{ko,irts,komsc,okmsc}). The detection of CIB
fluctuations at $\sim 0.7^{\rm o}$ from COBE DIRBE bands at 1.25
to 5 micron (\cite{ko}) shows a signal $\sim 2-3$ times that expected
from normal galaxy populations evolving via a present day IMF with
the observed star formation rates out to $z\sim 3-4$ (\cite{jk99}).
Recently, using combined 2MASS calibration data with long net
exposures, it became possible to remove resolved galaxies and
measure the near-IR CIB fluctuations spectrum from a few arcsec to
a few arcmin. We argued that this signal originated from faint
galaxies (K$\gsim 19$) at high $z$ (\cite{komsc,okmsc}), and is
significantly larger that what is expected from normal galaxy
populations. This suggests that the excess originates from still
earlier epochs.

\section{CIB flux and Population III}

The near-IR CIB signals detected in the 2MASS long integration data
(\cite{komsc,okmsc}) could come from the Population III stars
(\cite{ferrara,paper4,cooray,santos,salvaterra}). The argument can
be generalized and shown to be nearly model-independent provided
Population III stars were very massive.

Massive, metal-free stars will be dominated by radiation pressure
and would radiate close to the Eddington limit: $L \simeq L_{\rm
Edd} = \frac{4 \pi G m_p c}{\sigma_T} M \simeq 1.3 \times 10^{38}
M/M_\odot$ erg/sec, where $\sigma_T$ is the cross section due to
electron (Thompson) scattering. The energy spectrum for emission
from these stars will be nearly featureless and close to that of a
black body at $T \sim 10^{4.8-5}$K (\cite{schaefer,tumlinson}). Unlike
stars with higher metallicity, Population III stars are not expected
to have significant mass-loss during their lifetime (\cite{baraffe}).
If sufficiently massive ($\gsim 240 M_\odot$, see \cite{nosn})
such stars would also avoid SN explosions and collapse directly to
black holes. In this case their numbers could significantly exceed
the minimum that would be required to produce the metallicities
observed in Population II stars. The lifetime of these stars will
be independent of mass: $t_L\simeq \epsilon M c^2/L \simeq 3 \times
10^6$ years, where $\epsilon =0.007$ is the efficiency of the
hydrogen burning. These numbers are in good agreement with detailed
computations (\cite{schaefer}).

For a flat Universe the co-moving volume per unit solid angle
contained in the cosmic time interval $dt$ is $dV = c (1+z)^{-1}
d_L^2 dt$, where $d_L$ is the luminosity distance. The flux per
unit frequency interval from each of the Population III stars will
be $Lb_{\nu^\prime}(1+z)/4\pi d_L^2$, where $b_\nu$ is the fraction
of the total energy spectrum emitted per unit frequency and
$\nu^\prime = \nu(1+z)$ is the rest-frame frequency. (The Population
III SED is normalized so that $\int b_\nu d\nu=1$). The co-moving
mass density in these stars is $\int M n(M,t)dM =\Omega_{\rm baryon}
\frac{3H_0^2}{8 \pi G} f_*$, where $f_*$ is the fraction of the
total baryonic mass in the Universe locked in Population III stars
at time $t$. The net flux per unit frequency from a population of
such stars with mass function $n(M)dM$ is given by:
\begin{equation}
\frac{d}{dt} I_\nu = \frac{\int L n(M,t) dM}{4\pi d_L^2} \;(1+z) \; \langle{b_{\nu^\prime}}\rangle \;\frac{dV}{dt} \\= \frac{c}{4\pi} \langle b_{\nu^\prime} \rangle \langle \frac{L}{M} \rangle f_* \rho_{\rm baryonic}
\label{didt}
\end{equation}
Here $\langle b_\nu \rangle \equiv \int L n(M,t) b_\nu dM/\int L
n(M,t) dM$ denotes the mean Population III SED averaged over their
initial mass function and $\langle \frac{L}{M} \rangle \equiv \int
L n(M,t) dM/\int M n(M,t) dM$. For a Gaussian density field $f_*
\sim 5\times10^{-2} - 3\times 10^{-3}$, if on average Population
III formed in 2--3 sigma regions (see below). Provided that $L/M=$
constant, this result does not depend on the details of the initial
mass function of Population III stars (cf. Fig. 5 in ref. \cite{salvaterra}).
(Note that for the present day stellar populations their mass-to-light
ratio depends strongly on stellar mass and $\langle \frac{L}{M}
\rangle$ is much smaller than the Population III value of $4\pi G
m_p c/\sigma_T \simeq 3.3\times 10^4 L_\odot/M_\odot$ leading to
substantially smaller net fluxes). Assuming no significant mass
loss (\cite{baraffe}) during their lifetime $t_L$ for Population III
stars, these stars will produce CIB flux of amplitude:
\begin{equation}
\nu I_\nu = \frac{3}{8 \pi} \; \frac{1}{4\pi R_H^2} \; \frac{c^5}{G} \; \epsilon \Omega_{\rm baryon} \frac{1}{t_L} \int f_* \langle \nu^\prime b_{\nu^\prime} \rangle \frac{dt}{1+z} 
\label{cib_bol}
\end{equation}

Note that $c^5/G$ is the maximal luminosity that can be achieved
by gravitational processes and enters here because the nuclear
burning of stars evolves in gravitational equilibrium with the
(radiation) pressure. This result, eq. \ref{cib_bol}, has a simple
meaning illustrating its model-independence: the cumulative bolometric
flux produced is $L_{\rm max} = c^5/G \simeq 10^{26}L_\odot$
distributed over the surface of the Hubble radius, $4\pi R_H^2$,
times the model-dependent dimensionless factors. The spectral
distribution of the produced diffuse background will be determined
by $\langle\overline{b_\nu}\rangle\equiv \int f_*(1+z)^{-1} \langle
b_\nu \rangle dt/\int f_*(1+z)^{-1}dt$, which is the mean SED
averaged over the first star epoch (hereafter assumed to have duration $t_*$). The value of $L_{\rm
max}/4\pi R_H^2$ is $\simeq 3\times10^8 \; {\rm nW/m^2/sr}$ so even
with small values of $\Omega_{\rm baryon}, \epsilon, f_*$, the net
flux from Population III stars would be substantial. The K-band
roughly coincides with the Lyman limit (912 \AA) at $z_* \sim 20$
so much of the emitted flux would contribute today in the near IR
bands. If there was a population of massive black holes, as seems
likely to result from Population III evolution, their accretion
processes will emit radiation at a much higher efficiency of
$\epsilon \sim 0.1$ (and perhaps as high as $\epsilon \sim 0.4$
very close to the event horizon). At the same time, the K-band CIB
excess over the total contribution from the observed galaxy
populations ($\simeq 9 $\nwm2sr up to K$\sim 24$), is $\simeq 10-15
$\nwm2sr and can be completely accounted for by the contribution
from Population III stars at high redshifts.

The first stars formed in rare regions of the density field that
reached the turn-around over-density while the bulk of the matter
was still in a linear regime (density contrast $<1$). These regions
stopped expanding and began collapsing, forming the first generation
of stars if certain conditions are met. For a Gaussian density
field, as required for the CDM models, this would specify the value
of $f_*$ at each time. The power spectrum of the mass distribution
on large scales preserved the Harrison-Zeldovich power spectrum
index ($n\simeq 1$), but on scales less than the horizon scale at
the matter-radiation equality, the spectrum was modified by the
different growth-rates of sub- and super-horizon modes. This led
to a gradual decrease of $n$ toward the very small scale value of
$n \lsim -2$, suppressing the small scale power and leading to late
collapse and galaxy formation. In this picture, it is surprising
that the the first stars started forming at $z_* = 20^{+10}_{-9}$,
as indicated by the WMAP polarization measurements, but the most
straightforward explanation is that the earliest stars formed out
of very high peaks of the density field, leading to smaller $f_*$.

Assuming a spherical model for growth and collapse of density
fluctuations, any density fluctuation that reached the dimensionless
density contrast $\delta_{\rm col}=1.68$ will collapse. Consequently,
if $\sigma_M$ is the rms fluctuation at $z$ over a sphere containing
the total mass $M$, then $\eta \equiv \delta_{\rm col}/\sigma_M$
measures the numbers of standard deviations for that mass to collapse
at that redshift. Here, $\sigma_M$ is related to the underlying power
spectrum via $\sigma^2_M = \frac{1}{2\pi^2} \int P(k;z) W(kr_M)
k^2dk$ where $r_M=1.4 (M/10^{12}h^{-1}M_\odot)^{1/3}
(\Omega/0.3)^{-1/3}h^{-1}$Mpc is the co-moving linear scale containing
mass $M$, and $W(x)=(3j_1(x)/x)^2$. As they collapse, the primordial
clouds will quickly heat up to their virial temperature, $T_{\rm
vir}$, and only efficient cooling will enable isothermal collapse,
where gas pressure is smaller than gravity (or $T< T_{\rm vir}$).

\begin{figure}[h]
\centering
\leavevmode
\epsfxsize=0.8
\columnwidth
\epsfbox{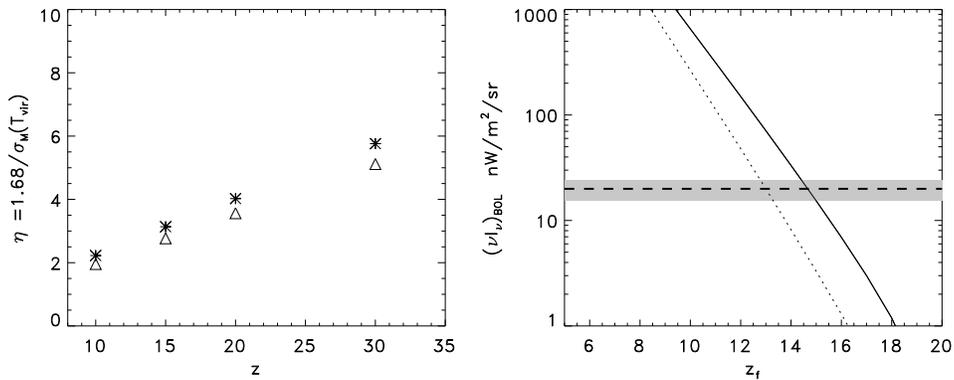}
\caption[]{
\scriptsize{
Left: The number of standard deviations that correspond to Population
III regions collapsing at $z$ for $\Lambda$CDM power spectrum of
the primordial density field. Triangles correspond to $T_{\rm vir}
= 400$K, asterisks to $T_{\rm vir} = 2\times 10^3$K. Right: cumulative
bolometric flux by Population III stars forming at $z_* =20$, and
lasting until $z_f$ shown on the horizontal axis. Solid line
corresponds to $T_{\rm vir} = 400$K and dotted line to $T_{\rm vir}
= 2\times 10^3$K. Thick dashed line shows the total CIB flux from
DIRBE measurements at J,K and L bands; the shaded region shows the
uncertainty. The values adopted were $22.9\pm 7.0$ KJy/sr in J band
(\cite{cambresy}) and $16.1\pm 4.4, 13.4 \pm 3.7$ KJy/sr in K, L
bands (\cite{wright}).
}
}
\label{eta}
\end{figure}

The details of what determines the collapse and formation of Population
III stars are not yet clear, but molecular cooling is critical
in the initial stages of metal-free collapse. Molecular cooling is
effective at $T\gsim 400$ K (\cite{santos}), but some simulations
suggest that rotational $H_2$ cooling can effectively dissipate
binding energy of the cloud only if $T > 2000$ K (\cite{miralda}).
We have evaluated the rms mass density fluctuation, $\sigma_M$,
for the $\Lambda$CDM model as a function of its virial temperature,
assuming a molecular weight of $\mu=1$. The left panel of Fig.
\ref{eta} shows the resulting $\eta$ vs $z$ for $T_{\rm vir} =$
400K and 2000K. For a Gaussian density field and $\eta >1$ the
value of $f_* \simeq {\rm erfc}(\eta/\sqrt{2})$. One can see that
Population III stars had to form out of 2 to 3 sigma fluctuations,
which for a Gaussian distribution correspond to $f_*$ varying
between $\sim 5\times 10^{-2}$ and $\sim 3\times 10^{-3}$. Departures
from spherical symmetry would accelerate the collapse,
decreasing $\eta$ and increasing $f_*$. Because $\eta$ is a decreasing
function of decreasing $z$, and $f_*$ is a (rapidly) increasing
function of decreasing $\eta$, the average $\overline{f_*}$ will
be dominated by the late times of Population III evolution.

The right panel of Fig. \ref{eta} shows that the total diffuse
background light produced during the first star era should be
substantial and measurable. Much of the diffuse flux produced by
the Population III stars will today be observable in the near-IR,
and could be responsible for the observed excess near-IR CIB. If
$z_f$ is large the net bolometric flux from Population III is
smaller, but because of the redshift effects a larger fraction of
the bolometric flux will be observed today in the near-IR bands.
Furthermore, the near-IR part of the CIB emission from Population
III would be increased because much of the shorter wavelengths
emission gets absorbed and re-emitted at these bands by Lyman and
free-free emission from the ionized nebulae around Population III
stars (\cite{santos}). Dashed line shows the total CIB flux evaluated
from the DIRBE 1.25 to 3.5 $\mu$m bands. Taken at face value that
would imply that the first star era lasted until $z_f \lsim15$.
However, Population III stars only contribute a fraction of the
observed CIB.

\section{CIB anisotropies from Population III}

The contribution of Population III stars to the CIB will possess
a spatial structure that is related to that of density fluctuations
from which the stars formed. At particular spatial scales and
wavelengths it may prove easier to distinguish the Population III
contribution from the various foregrounds through the spatial
structure rather than the measurement of the mean (isotropic)
intensity.

In the limit of small angles the 2-dimensional angular power
spectrum, $P_2(q)$, of the CIB fluctuations that arise from
clustering of Population III systems with the 3-dimensional power
spectrum $P_3(k)$ is given by the modified Limber equation (\cite{ko}):
\begin{equation}
P_2(q) = \frac{1}{c} \int \left( \frac{d I_{\nu^\prime}}{dt}\right)^2 \frac{P_3(qd_A^{-1};z)}{d_A^2} dt 
\label{limber}
\end{equation}
where $d_A$ is the comoving angular diameter distance. The power spectrum $P_3$ is in turn related to the underlying
$\Lambda$CDM model power spectrum and its evolution and growth. On
larger angular scales ($\theta \gsim $ a few arcmin) the power
spectrum is in the linear regime and we evaluate the growth rate
of fluctuations using linear theory for the $\Lambda$-dominated
cosmology.

\begin{figure}[h]
\centering
\leavevmode
\epsfxsize=0.5
\columnwidth
\epsfbox{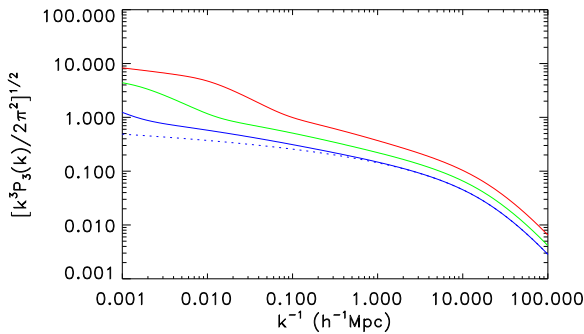}
\caption[]{
\scriptsize{
$\Lambda$CDM density field at $z=6,10,15$ (red, green and blue).
Dotted line corresponds to linearly evolving density field at
$z=15$.
}
}
\label{p3}
\end{figure}

On sufficiently small scales the primordial power spectrum of the
density field will be modified by non-linear gravitational effects.
Fig. \ref{p3} shows the density field for the $\Lambda$CDM model
at various high redshifts. The linear power spectrum was taken to
be that of the $\Lambda$CDM for the WMAP cosmological parameters.
The non-linear evolution of the density field was modeled analytically
using the approximation from ref. (\cite{peacock}). For reference 1 arcmin subtends the comoving scale of $\sim 1.5
h^{-1}$Mpc at $z$ between 6 and 15 for WMAP cosmological parameters and a dark energy
equation of state with $w=-1$. The dotted line shows the linearly
evolved density field at $z=15$. On sub-arcminute scales the density
field is in the quasi-linear to non-linear regime (density contrast
$\gsim 0.2$) for the $\Lambda$CDM model. There the spectrum due to
clustering evolution was modified significantly and the fluctuations
amplitude has increased from its primordial value, especially since
the effective spectral index on these scales for $\Lambda$CDM model
was $n \lsim -2$.

To within a factor of order unity, the square of the fractional
fluctuation of the CIB on angular scale $\simeq \pi/q$ is $\delta_{\rm
CIB}^2 = \langle (\delta I_\nu)^2\rangle/I_\nu^2 \simeq I_\nu^{-2}
q^2P_2(q)/2\pi$. The meaning of eq. \ref{limber} can be demonstrated
by assuming $dI_{\nu^\prime}/dt$=constant during the Population
III phase of duration $t_*$. In this case the fractional fluctuation
due to clustering of early star systems becomes: \begin{equation}
\delta_{\rm CIB}^2 = \frac{\pi}{t_*} \int \Delta^2(qd_A^{-1};z) dt
\label{spec_cib} \end{equation} Here $\Delta (k) = [\frac{1}{2\pi^2}
\frac{k^2 P_3(k)}{ct_*}]^{1/2}$ is the fluctuation in number of
sources within a volume $k^{-2} ct_*$. In other words, the fractional
fluctuation on angular scale $\pi/q$ in the CIB from Population
III stars is given by the average value of the r.m.s. fluctuation
from Population III spatial clustering over a cylinder of length
$ct_*$ and diameter $\sim k^{-1}$. At $z=20$ the Universe is $\sim
2 \times 10^8$ years old which is much larger than the age of the
individual Population III stars, $t_L \sim 3 \times 10^6$ years.
If the first stars lasted for only one generation forming at $z_*$,
the relative CIB fluctuations will be $\delta_{\rm CIB} \sim
\sqrt{\pi}\Delta(qd_A^{-1};z)|_{z_*}$.

Several points are worth noting about eq. \ref{limber}: Firstly, the
value of $\Delta(k)$ is inversely proportional to $\sqrt{t_*}$ and
thus measures the duration of the first star era. The density
perturbations grow with decreasing $z$, so most of the contribution
to the integral in eq.\ \ref{limber} comes from the low end of $z$.
The overall dependence on $t_*$ wins out and the longer the Population
III phase, the smaller are the relative fluctuations of the CIB
from them. Secondly, in the Harrison-Zeldovich regime of the power spectrum,
$P_3 \propto k$, one would have $\delta_{\rm CIB} \propto q^{1.5}$.
Finally, the transition to the Harrison-Zeldovich regime occurs in the
linear regime at all relevant redshifts and happens at the co-moving
scale equal to the horizon scale at the matter-radiation equality.
This feature would occur
at different angular scales as the redshift of the Population III era
decreases. All this would allow probes of the epoch of the first
stars, its duration, and the primordial power spectrum at high
redshifts on scales that are currently not probed well. Interestingly,
a short duration for the first stars will lead to smaller CIB flux,
but larger relative fluctuations and vice versa.

In addition to the small angular scale increase due to non-linear
gravitational evolution, the fluctuations in the distribution of
Population III systems will be amplified because, as Fig. \ref{eta}
shows, these systems had to form out of rare peaks of the primordial
density field. Consequently, their 2-point correlation function
will be amplified (\cite{kaiser,szalay,politzer,ak1987,ak1998}) over
that given by the $\Lambda$CDM power spectrum, $\xi =\frac{1}{2\pi^2}
\int P_3(k)j_0(kr)k^2 dk$. Adopting a Press-Schechter prescription
for the formation of collapsed objects (\cite{press-schechter}) from a
correlated density field leads to the following expression for the
amplification of the 2-point correlation function (\cite{ak1998}):
\begin{equation}
\xi_\eta = \xi \sum_{m=0}^\infty \frac{\eta^{2m}}{ m!\;\delta_{\rm
col}^{2m} }\;\xi^m \; \left[ \frac{H^2_{m+1}(\eta/\sqrt{2})}{2\eta^2}
+\frac{H^2_{m+2}(\eta/\sqrt{2})}{4(m+1)\delta_{\rm col}^2} \right]
\end{equation}
where $H_n$ are Hermite polynomials. At large values of $\eta$ this
reduces to $\xi_\eta \simeq \exp(\eta^4 \xi/\delta_{\rm col}^2)-1$.
At small values of $\xi \ll 1$ (and sufficiently large angular
scales) the amplification factor is linear and equal to $\simeq
\eta^4/\delta_{\rm col}^2$. At very small scales the amplification
is non-linear in $\xi$ and will exceed the linear amplification
value leading to {\it larger} CIB fluctuations. We computed the
resultant CIB fluctuations using the linear bias amplification
approximation, but note that this gives a {\it lower} limit on the
resultant CIB fluctuations.

\begin{figure}[h]
\centering
\leavevmode
\epsfxsize=0.4
\columnwidth
\epsfbox{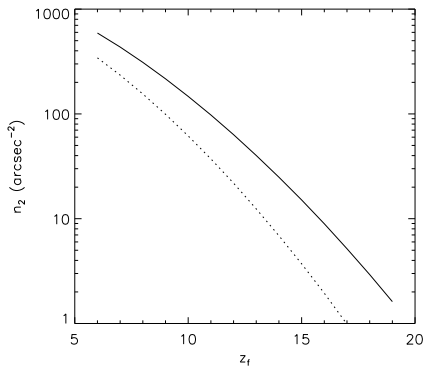}
\caption[]{
\scriptsize{
Mean projected surface density of Population III systems along the
line of sight, plotted vs the redshift marking the end of the
Population III era. Solid and dotted lines correspond to primordial
clouds containing Population III with $T_{\rm vir}>$ 400 K and 2000
K respectively.
}
}
\label{shotnoise}
\end{figure}

There will also be shot-noise fluctuations due to individual
Population III systems entering the beam. The relative magnitude
of these fluctuations will be $N_{\rm beam}^{-1/2}$, where $N_{\rm
beam}$ is the number of the Population III systems within the beam.
This component will be important at very small angular scales,
where $N_{\rm beam} \sim 1 $, and will contribute to the power
spectrum: $P_{\rm SN} = 1/n_2$, where $n_2 = c \int n_3
d_L^2 (1+z)^{-1} dt$ is the projected angular number density of
Population III systems. Detection of the shot-noise component
in the CIB power spectrum at small angular scales will give a direct
measure of both the duration of the first star era and constrain
the makeup and masses of the Population III systems. The
three-dimensional number density can be evaluated from the
Press-Schechter (\cite{press-schechter}) formalism, but in the
appropriate limits ($\eta > 1$) it becomes $n_3 \sim \rho_{\rm
baryon}M^{-1}{\rm erfc}(\eta(z)/\sqrt{2})$. Fig. \ref{shotnoise}
shows that the mean projected surface density of Population III
systems $n_2$ for $T_{\rm vir}>$ 400 K and 2000 K. One can see that
unless the Population III era was very short, the shot noise
correction to the CIB would be small on scales greater than a few
arcseconds.

Essentially all of the flux emitted at rest wavelength less than 912
\AA\ will be absorbed by the Lyman continuum absorption in the
surrounding medium (\cite{yoshii}) and the contribution to the CIB
will be cut off at $\lambda < 912 (1+z_f)$ \AA . At longer wavelengths
the flux, including emission from the nebulae around each star,
will propagate without significant attenuation. Population III
stars emit as black bodies at $\log T \simeq 4.8-5$ (\cite{schaefer})
and the Lyman limit at $z\sim 20$ is shifted to $\sim 2 \mu$m in
the observer's frame. The 2MASS J band filter contains emission
out to 1.4 \um\, and represents the shortest wavelength where excess
in the CIB over that from normal galaxies has been measured
(\cite{cambresy,komsc}). If the measured excesses in CIB fluctuations
(\cite{komsc,ko}) and isotropic component (\cite{cambresy}) at J band
are indeed attributable to Population III, their era must have
lasted until $z \lsim 14$.

\begin{figure}[h]
\centering
\leavevmode
\epsfxsize=1.
\columnwidth
\epsfbox{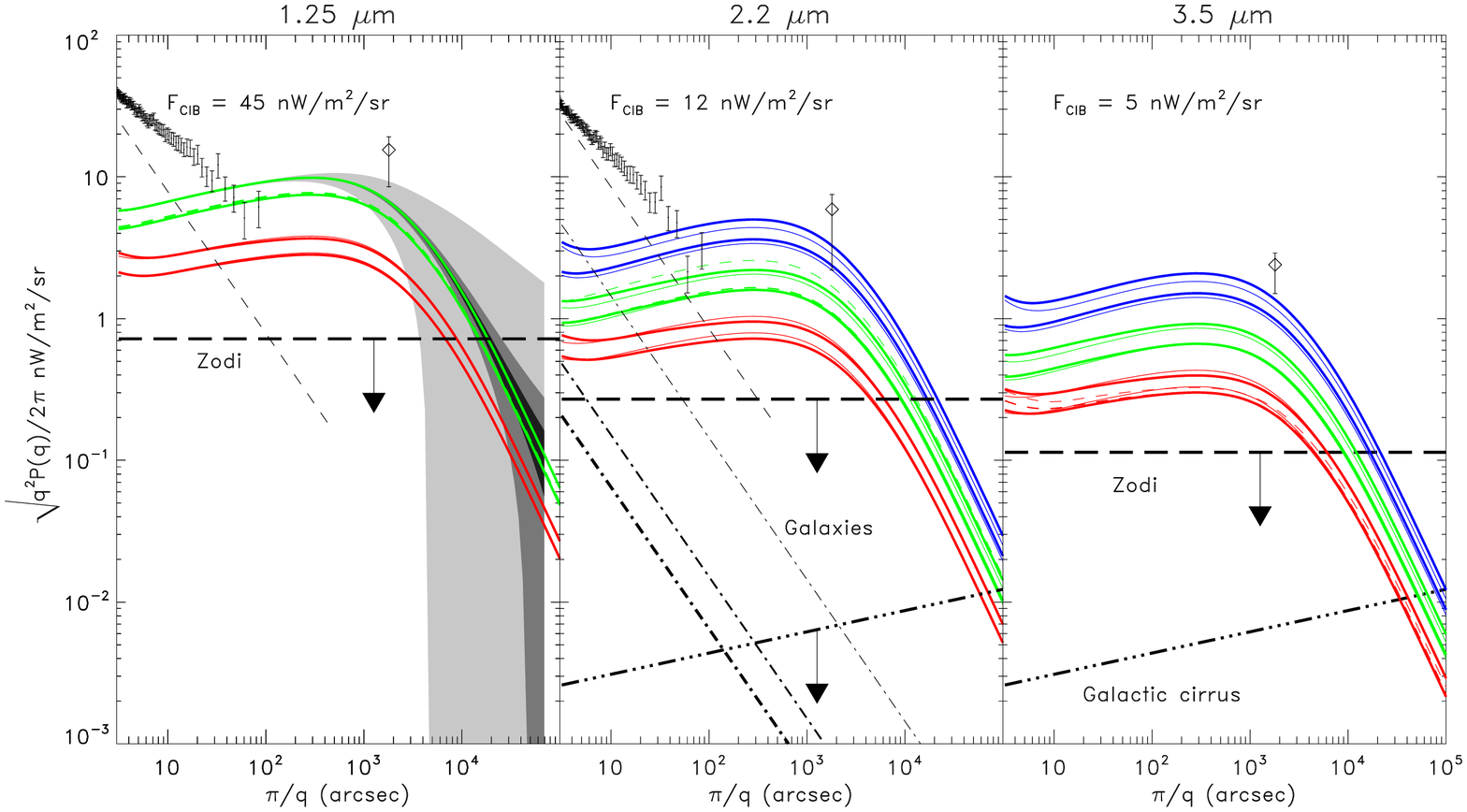}
\caption[]{
\scriptsize{
CIB fluctuations, $\sqrt{q^2P_2/2\pi}$, from Population III stars
forming at $z_*=20$. Continuous blue, green and red lines
correspond to $z_f = 15, 10, 6$ respectively. Thin colored solid lines
correspond to the case of constant CIB flux rate production and $b_\nu$ = const. Thick
colored solid lines correspond to $dI_\nu/dt \propto f_* b_{\nu^\prime}$
with $b_\nu \propto \nu^2$. Lower lines correspond to $T_{\rm
vir}=400$K, and top lines to $T_{\rm vir}=2000$K. Colored dashed lines show the CIB fluctuations for more realistic SED from Santos et al (2002) which include the Lyman-$\alpha$ and free-free emission from the surrounding nebula. They mostly overlap with the other SED models showing the relative model-independence of the results. Thin and thick dashed colored lines correspond to SED from Fig. 3 (completely opaque nebula to ionizing photons) and Fig. 5 (completely transparent nebula). Thin dashed lines
show atmospheric fluctuations from ground based 2MASS measurements
after 1 hour of exposure. Thick long dashes denote the upper limit
on zodiacal light fluctuations from \cite{abraham} scaled to the
corresponding band. Thick dashed-triple-dotted line denotes cirrus
fluctuations: these are upper limits at 2.2 micron and an estimate
from \cite{kiss} at 3.5 micron. In the middle panel dot-dashed
lines correspond to shot noise from galaxies fainter than K=24
(thickest), K=23 and K=20 magnitudes (thin). The K-band faint galaxy
counts data used were taken from \cite{totani}. Diamonds with error
bars show the CIB fluctuations at $\sim 0.7^\circ$ from the COBE
DIRBE fluctuations analysis (\cite{ko}). Note that because of the
large DIRBE beam, the results include contribution from all galaxies
as well as other sources such as Population III. Dots with error bars show the K band CIB
fluctuation from deep integration 2MASS data (\cite{komsc}. The
2MASS data shown in the figure were taken for the the patches for
which galaxies were removed brighter than $K \simeq 19^m$ (\cite{okmsc}).
The cosmic variance 1-sigma uncertainty
is shown with shaded regions in the left panel. The darkest shade
corresponds to a total of 1,000 deg$^2$ observed, the intermediate 
shade corresponds to a total of 100 deg$^2$ and the lightest shade corresponds to 1 deg$^2$.
}
}
\label{del_cib}
\end{figure}

Fig. \ref{del_cib} shows the resultant CIB fluctuations from
Population III stars at 1.25, 2.2 and 3.5 micron or J, K and L
photometric bands. The assumed CIB fluxes from Population III are
listed near the tops of the panels. At 2.2 micron the galaxies
contribute around 9 \nwm2sr whereas the net CIB flux is around
20-25 \nwm2sr ; hence the assumed excess of 10-15 \nwm2sr is
attributed to the Population III stars for the purposes of this
discussion and is taken to be 12 \nwm2sr in the figure. The situation
is similar at 3.5 micron (\cite{dwek,gorjian}) and the excess was
taken to be 5 \nwm2sr (\cite{wright}). The excess in the J band over
the total flux from normal galaxies was taken from (\cite{cambresy}).
The clustered component of the CIB fluctuation displayed in the
figure can be rescaled to other CIB fluxes via eq. \ref{spec_cib}.
Thin colored lines assume constant rate of production of the
Population III fraction of the CIB and the flat SED with $b_\nu=$ constant. All Population III stars were
assumed to start forming at $z_* = 20$, but different colored lines
correspond to different values of $z_f$, the redshift of the end
of the first star era. Thick colored lines correspond
to a Population III contribution produced by black-body emitters
in the Rayleigh-Jeans regime ($b_\nu \propto \nu^2$) with the rate
of flux production $\propto f_*$. We also computed CIB fluctuations with the rate of flux production $\propto f_*$ for more realistic SED from Santos et al (2002) which include the Lyman-$\alpha$ and free-free emission from the surrounding nebula. Thin and thick dashed colored lines correspond to two possible extremes: SEDs from their Fig. 3 (completely opaque nebula to ionizing photons) and Fig. 5 (completely transparent nebula). Because they do not differ significantly from the more simple SEDs the lines are shown only for some values of $z_f$ to avoid overcrowding in the Figure.  The graphs show the relative model-independence of the relative CIB fluctuations on the SED for a fixed total CIB flux; the latter contains the bulk of information on the details of Population III emission distribution between various NIR bands. For a given Population III CIB isotropic
component, the fluctuations are largest for large values of $z_f$
because the duration of the first star era is the smallest. Because
of the Lyman continuum absorption there would be no appreciable
CIB fluctuations at J band if $z_f \gsim 14$ (no blue lines in the
left panel) and the contribution to the other scenarios will come
from $z \sim 13$, rather than $z=20$ when the first star era was
assumed to start. This would be broadly consistent with the limits on sky brightness fluctuations at 4 \um\ from Xu et al (2003).

Both DIRBE and 2MASS data indicate CIB anisotropies at amplitudes
larger than the contribution from normal galaxy populations
(\cite{ko,komsc}) and are consistent with significant contributions
due to Population III. However, because the measured signal contains
contributions from remaining galaxies (all galaxies for DIRBE and
$K\geq 19$ galaxies for 2MASS), it is difficult to isolate the
contribution from the Population III stars.

\section{Prospects for measuring CIB from Population III}

 Population III stars, if massive, should have left a unique and
measurable signature in the near-IR CIB anisotropies over angular
scales from $\sim 1$ arcminute to several degrees as Fig. \ref{del_cib}
shows. As is the case for measurement of the mean isotropic CIB,
detection of these fluctuations depends on the identification and
removal of various foreground emission (and noise) contributions:
atmosphere (for ground-based measurements), zodiacal light from
the Solar system, Galactic cirrus emission, and instrument noise.

The dashed line in Fig. \ref{del_cib} shows the small scale atmospheric
fluctuations at 2 \um\ after one hour of integration of ground
based measurement (\cite{okmsc}). At large angular scales atmospheric gradients will become important, making measurements there even more difficult. Atmospheric effects can be highly
variable on a wide range of time scales, yet they can be completely
avoided with space--based experiments. Zodiacal emission from
interplanetary dust (IPD) is the brightest foreground at most IR
wavelengths over most of the sky. There are some structures in this
emission associated with particular asteroid families, comets, and
an earth--resonant ring, but these structures tend to be confined
to low ecliptic latitudes or otherwise localized. The main IPD
cloud is generally modeled with a smooth density distribution.
Observationally, intensity fluctuations of the main IPD cloud have
been limited to $<0.2$\% at 25 $\mu$m (\cite{abraham}). Extrapolating
this limit to other wavelengths using the mean high--latitude
zodiacal light spectrum yields the limits shown in Figure \ref{del_cib}.
Because the Earth is moving with respect to (i.e. orbiting within) the
IPD cloud, the zodiacal light varies over time. Likewise, any
zodiacal light fluctuations will not remain fixed in celestial
coordinates. Therefore repeated observations of a field on timescales
of weeks to months should be able to distinguish and reject any
zodiacal light fluctuations from the invariant Galactic and CIB
fluctuations.

Intensity fluctuations of the Galactic foregrounds are perhaps the
most difficult to distinguish from those of the CIB. Stellar emission
may exhibit structure from binaries, clusters and associations,
and from large scale tidal streams ripped from past and present
dwarf galaxy satellites of the Milky Way. At mid- to far-IR
wavelengths, stellar emission is minimized by virtue of being far
out on the Rayleigh--Jeans tail of the stellar spectrum (apart from
certain rare classes of dusty stars). At near-IR wavelengths stellar
emission is important, but with sufficient sensitivity and angular
resolution most Galactic stellar emission, and related structure,
can be resolved and removed. IR emission from the ISM (cirrus) is
intrinsically diffuse and cannot be resolved. Cirrus emission is
known to extend to wavelengths as short as 3 $\mu$m. Statistically,
the structure of the cirrus emission can be modeled with power--law
distributions. Using the mean cirrus spectrum, measurements made
in the far-IR can be scaled to 3.5 $\mu$m, providing the estimated
fluctuation contribution from cirrus that is shown in Figure
\ref{del_cib}. The extrapolation to shorter wavelengths is highly
uncertain, because cirrus (diffuse ISM) emission has not been
detected at these wavelengths, and the effects of extinction may
become more significant than those of emission, but the cirrus
contribution is expected to be several orders of magnitude lower
than the CIB fluctuations.

The current detections of CIB anisotropies from DIRBE and 2MASS
are above all these foregrounds. The level of anisotropy compares
favorably with theoretical expectations and may indeed contain a
significant signal from Population III stars. However, the anisotropy
may also contain contributions from galaxies made up of Populations
II and I stars, which may be significant and uncertain at high $z$
because of unknown evolution effects. Hence it is important to be
able to remove galaxy contribution in future measurements. Galaxies
with normal stellar populations fainter than $K\sim 24$ contribute
little to the CIB. At fainter magnitudes (and larger redshifts) the emission is
likely to be dominated by Population III stellar populations and
protogalaxies. Because each unit mass of Population III emits a
factor of $\sim 10^5$ more light that the normal (Population I,
II) stellar populations, the diffuse flux from (massive) Population
III stars would dominate emission from faint normal galaxies (say
$K\gsim 24$). The figure shows that CIB fluctuations from Population
III would be the dominant source of diffuse light fluctuations on
arcminute and degree scales even if Population III stars epoch was
briefer, and their diffuse flux smaller, than the current CIB
numbers suggest. Their angular power spectrum should be very
different from other sources of diffuse emission and its measurement
thus presents a way to actually discover Population III and measure
the duration of their era and their spatial distribution. The latter
would provide direct information on primordial power spectrum on
scales and at epochs that are not easily attainable by conventional
surveys. This measurement would be imperative to make and is feasible
with the present day space technology.

\section{Observational requirements}

 There are several ways in which observations could verify the
analysis of this paper. The James Webb Space Telescope (JWST) will
be capable of resolving individual Population III objects in their
supernova phase, or as young black holes accreting substantial mass
flows. Alternatively, if the Population III objects are sufficiently
clustered in space and time, the aggregations would resemble
proto-galaxies and could be recognized. Observations of these
objects are among the foremost objectives of the JWST.

The following would be the optimal set of parameters for an
observation designed to map the contribution of the first star era
to CIB anisotropies: 1) In order to make certain that the signal
is not contaminated by distant galaxies with normal stellar
populations, one would need to conduct a deep enough survey in
order to identify and eliminate normal galaxies from the field. In
practice, as Fig. \ref{cib_dc} indicates, going to K$\simeq 24$
would be sufficient. With the current technology this
would take $\sim 1.5$ hours for a 1 meter space telescope. 2) A direct
signature of Population III signal is that there should be no CIB
fluctuations at wavelengths $\leq 912 (1+z_f)^{-1}$ \AA. That
spectral drop would provide an indication of the epoch corresponding
to the end of first star era. At longer wavelengths the spectral
energy distribution of the CIB fluctuations would probe the history
of energy emission and its re-distribution by the nearby gas during
the first star era. At wavelengths $\gsim 10 \mic$ zodiacal light
fluctuations may become dominant. 3) On angular scales from a few arcminutes to $\sim 10^\circ$ Population III would produce CIB anisotropies with a distinct and measurable angular spectrum, but its measurement will be limited by the cosmic or sampling variance. If the power spectrum is determined from fraction $f_{\rm sky}$ of the sky by sampling in concentric rings of radius $q\sim \pi/\theta$ and width $\Delta q$, the relative uncertainty on $P_2(q)$ will be $\sim (2\pi)^{-1} {\theta\over180^\circ} \sqrt{q\over\Delta q}\;f_{\rm sky}^{-1/2} $. In order to get accurate and independent measurements, it is necessary to have small $\Delta q/q \sim 0.05-0.1$; thus for reliable results on scales up to $\theta$ one would need to cover area a few times larger. Shaded regions in the left panel of Fig. \ref{del_cib} show the one sigma cosmic variance uncertainty with $\Delta q/q = 0.05$ for a total of 1, 100 and 1,000 deg$^2$ areas covered. In order to get reliable results on scales up to $\theta\gsim 1^\circ$ one would need to observe areas of $\gsim 10^\circ$ across.

\vspace{0.2cm}
\begin{figure}[h]
\centering
\leavevmode
\epsfxsize=1.
\columnwidth
\epsfbox{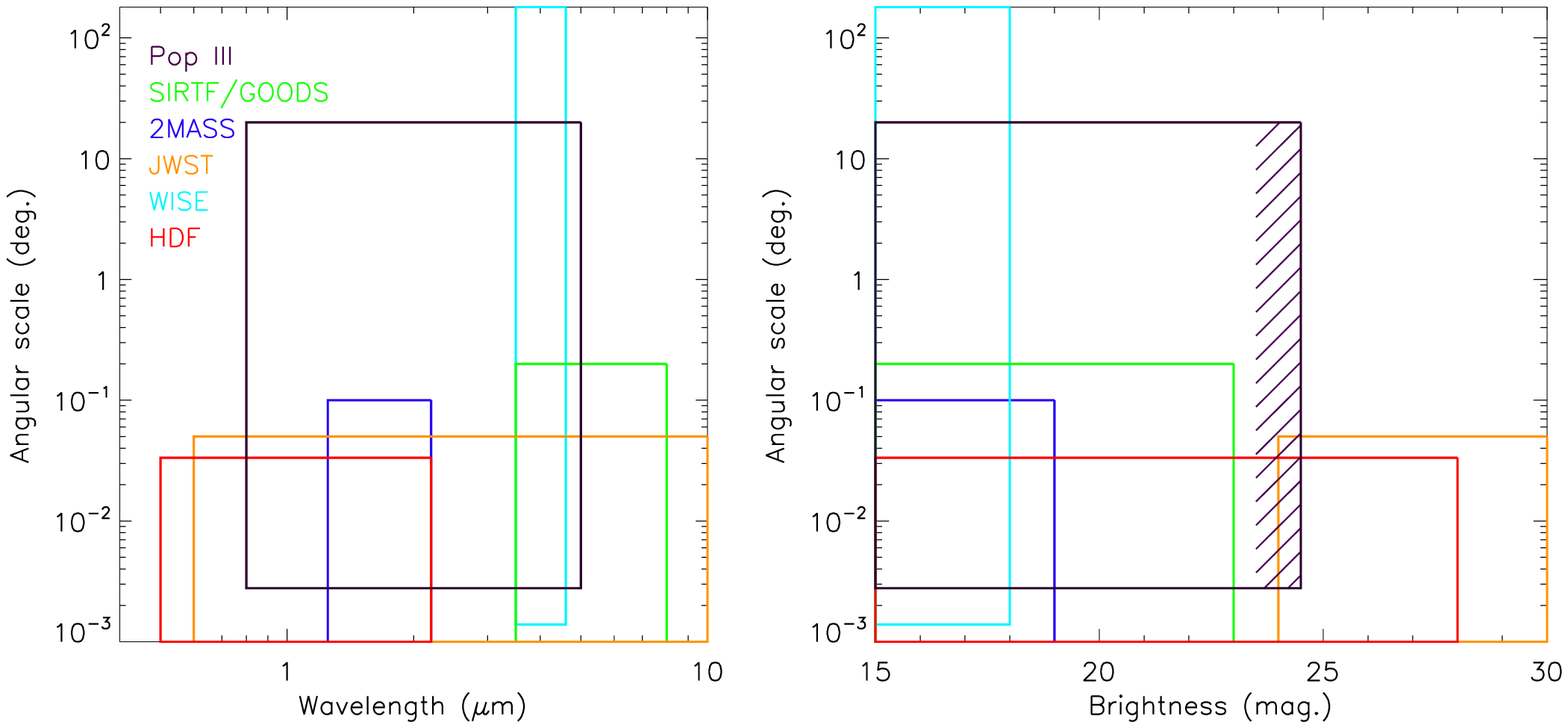}
\caption[]{
\scriptsize{Black rectangular box shows the regions of the parameters
for a mission to measure the CIB anisotropies from Population III
thereby detecting the first star era. Cross-hatched region in the right panel displays the magnitude range where Population III stars are expected to dominate CIB anisotropies.Other colors correspond to
the currently existing datasets or planned NASA missions useful
for this purpose: red is the Hubble Deep Field from HST
(http://www.stsci.edu/ftp/science/hdf/hdf.html), dark blue is for
the deep 2MASS data (\cite{komsc,okmsc}), green is the NASA SIRTF
GOODS project (http://www.stsci.edu/ftp/science/goods/), light blue
corresponds to the planned NASA WISE MIDEX mission
(http://www.astro.ucla.edu/~wright/WISE), and orange color corresponds
to the field-of-view of the NASA James Webb Space Telescope planned
for the end of the decade (http://jwst.gsfc.nasa.gov).
}
}
\label{pop3e}
\end{figure}

The black box in Fig. \ref{pop3e} shows what parameters the ideal
space mission should have for such a measurement with cross-hatched region in the right panel displaying the magnitude range where Population III stars are expected to dominate CIB anisotropies. For comparison
other currently operating or planned NASA missions in the near-IR
are also shown. Some of the datasets overlap with the required
parameter space, but a large part of the desired parameter space
can not be reached with these other data sets.

Direct observations of the CIB fluctuations with the needed fidelity
and angular range is not easy. The foreground objects (stars and
low-redshift galaxies, and the interstellar medium) are bright and
abundant. However, they do not have the same spatial and spectral
distributions as the predicted CIB fluctuations. The first
observational requirement is to measure the fluctuations as a
function of spatial scale and wavelength, with sufficient corroborating
evidence to show that the fluctuations are not the residual of some
foreground process.

The predicted fluctuations turn down at angular scales greater
than subtended by the horizon at decoupling seen at the first star
era, and this must be verified. Masking out contributions from
individual galaxies requires angular resolutions of the order of
1 arcsec and adequate depth to remove galaxies of K $\lsim 23-24$.
Observing fluctuations on the order of 3 degrees in size with
reasonable resolution of the spatial scale and reasonable
signal-to-noise ratio requires a map of the order of 1000 deg$^2$
in size, about 1/40 of the whole sky. Taken at face value, we need
a map of the order of $\simeq 10^{10}$ pixels, to be built up as
a mosaic of exposures with individual detectors and telescope
pointings.

There are many technical challenges associated with such a project.
Ideally the observations would be taken in space, to avoid the
limitations of atmospheric noise, and to enable observations at
the full range of wavelengths from 1 to 5 {\um}. In addition, we
would require software to stitch together exposures from different
detectors, taken at different times with different telescope
orientations, and different amounts of zodiacal light and cosmic
ray hits on the detectors, while preserving accurate calibration.
It is essential to preserve large-scale photometric accuracy to
enable measurements of fluctuations on scales larger than a single
field of view (\cite{fixsen}). To understand and eliminate
the effects of interstellar dust and PAH clouds, it will be necessary
to provide photometric bands that maximize the detectability of
these sources of interference. In addition, observations over wide
ranges of foreground brightness would enable recognition of unexpected
effects.
Such a project would produce a large database of space observations
of low-contrast, extended features, and could lead to unexpected
discoveries.

To summarize, the measurement of CIB fluctuations from the Population III star era is imperative to make and is feasible
with the present day space technology.



\begin{thebibliography}{3}
\bibitem [Abel et al 2002]{abel}{Abel, T. et al 2002, Science, 295, 93}
\bibitem [Abraham et al 1997]{abraham}{\'Abrah\'am, P., Leinert, Ch., \& Lemke, D. 1997, A\&A, 382, 702-705}
\bibitem [Baraffe et al 2001]{baraffe}{Baraffe, L, Heger, A. \& Woosley, S.E. 2001, Ap.J.,550,890}
\bibitem [Bennett et al 2003]{wmap}{Bennett, C. L. et al 2003,Ap.J.Suppl.,148,1}
\bibitem [Bond et al 1984]{nosn}{Bond, J.R., Arnett, W.D. \& Carr, B.J. 1984,Ap.J.,280,825}
\bibitem [Bond et al 1986]{bond}{Bond, J.R., Carr, B.J. \& Hogan, C. Ap.J., 1986,306,428}
\bibitem [Bromm et al 1999]{bromm}{Bromm, V. et al 1999, Ap.J., 527, L5}
\bibitem [Cambresy et al 2001]{cambresy}{Cambresy, L. et al 2001,Ap.J.,555,563}
\bibitem [Cooray, A. et al 2003]{cooray}{Cooray, A. et al 2003, astro-ph/0308407}
\bibitem [Dwek \& Arendt 1998]{dwek}{Dwek, E. \& Arendt, R. 1998, Ap.J.,508,L9}
\bibitem [Fixsen et al 2000]{fixsen}{Fixsen, D. J., Moseley, S. H., Arendt, R. G.  2000, Ap.J.Supp., 128, 651}
\bibitem [Gardner 1996]{gardner}{Gardner, J.P. 1996,in ``Unveiling the cosmic 
infrared background", ed. E. Dwek, p. 127}
\bibitem [Gorjian et al 2001]{gorjian}{Gorjian, V. et al 2001,536,550}
\bibitem [Hauser et al 1998]{hauser}{Hauser, M. et al 1998, Ap.J.,508,25}
\bibitem [Hauser \& Dwek 2001]{cibreview}{Hauser, M. \& Dwek, E. 2001, Ann Rev A A, 39, 249}
\bibitem [Jensen \& Szalay 1986]{szalay}{Jensen, L. \& Szalay, A., 1986, Ap.J.,305,L5}
\bibitem [Jimenez \& Kashlinsky 1999]{jk99}{Jimenez, R. \& Kashlinsky, A. 1999, Ap.J.,511,16}
\bibitem [Kaiser 1984]{kaiser}{Kaiser, N. 1984, Ap.J.,284,L9}
\bibitem [Kashlinsky 1987]{ak1987}{Kashlinsky, A. 1987, Ap.J.,317,19} 
\bibitem [Kashlinsky 1998]{ak1998}{Kashlinsky, A. 1998, Ap.J.,492,1}
\bibitem [Kashlinsky et al 1996]{paper1}{Kashlinsky, A., Mather, J.C., Odenwald, S. \& Hauser, M. 1996, Ap.J.,470,681}
\bibitem [Kashlinsky et al 1999]{paper4}{Kashlinsky, A., Mather, J.C., Odenwald, S. 1999, preprint}
\bibitem [Kashlinsky \& Odenwald 2000a]{ko}{Kashlinsky, A. \& Odenwald, S. 2000a, Ap.J.,528,74}
\bibitem [Kashlinsky \& Odenwald 2000b]{koscience}{Kashlinsky, A. \& Odenwald, S. 2000b, Science,289,246}
\bibitem [Kashlinsky et al 2002]{komsc}{Kashlinsky, A., Odenwald, S., Mather, J.C., Skrutskie, M. 2002, Ap.J,579,L53}
\bibitem [Kiss et al 2003]{kiss}{Kiss, Cs., et al 2003, A\&A, 399, 177-185}
\bibitem [Kogut et al 2003]{kogut}{Kogut, A. et al 2003, Ap.J.Suppl.,148,161}
\bibitem [Madau \& Pozzetti 2000]{madau}{Madau, P. \& Pozzetti, L. 2000,MNRAS,312,L9}
\bibitem [Magliochetti et al 2003]{ferrara}{Magliochetti, M., Salvaterra, R., Ferrara, A. 2003, MNRAS,342,L25}
\bibitem [Matsumoto et al 2000]{irts}{Matsumoto, M. et al 2000,in ``ISO surveys of a dusty Universe", eds. Lemke, D. et al. p.96}
\bibitem [Miralda-Escude 2003]{miralda}{Miralda-Escude, J. 2003, Science, 300, 1904}
\bibitem [Odenwald et al 2003]{okmsc}{Odenwald, S., Kashlinsky, A., Mather, J.C., Skrutskie, M. 2003, Ap.J,583,535}
\bibitem [Peacock \& Dodds 1996]{peacock}{Peacock, J.A. \& Dodds, S.J., 1996, MNRAS,280,L19}
\bibitem [Politzer \& Wise 1984]{politzer}{Politzer, H.D. \& Wise, M.B. 1984, Ap.J.,285,L1}
\bibitem [Press \& Schechter 1974]{press-schechter}{Press, W. \& Schechter, P. 1974, Ap.J., 187, 425}
\bibitem [Rees 1978]{mjr}{Rees, M.J. 1978, Nature,275,35}
\bibitem [Salvaterra \& Ferrara 2003]{salvaterra}{Salvaterra, R. \& Ferrara, A. 2003, MNRAS,339,973}
\bibitem [Santos et al 2002]{santos}{Santos, M.R., Bromm, V., Kamionkowski, M. 2002,MNRAS,336,1082}
\bibitem [Schaerer 2002]{schaefer}{Schaerer, D. 2002, Aston. Astrophys., 382,28}
\bibitem [Schneider et al 2002]{ostriker}{Schneider, S. et al 2002,Ap.J.,571,30}
\bibitem [Totani et al 2001]{totani}{Totani, T. et al 2001,Ap.J.,559,592}
\bibitem [Tumlinson et al 2003]{tumlinson}{Tumlinson, J., Shull, J.M., \& Venkatesan, A. 2003,Ap.J.,584,608}
\bibitem [Wright 2003]{wright}{Wright, E.L. 2003,astro-ph/0306058}
\bibitem [Xu et al 2002]{nite}{Xu, J. et al 2002, Ap.J., 580, 653}
\bibitem [Yoshii \& Peterson 1994]{yoshii}{Yoshii, Y. \& Peterson, B. 1994, Ap.J.,436,551}
\end{thebibliography}
\end{document}